\documentclass[fleqn,10pt]{wlscirep}
\usepackage{physics}
\usepackage{amsmath}
\usepackage{amssymb}
\usepackage{multirow, array}

\title{Interference effects in quantum-optical coherence tomography using spectrally engineered photon pairs}

\author[1]{Pablo Yepiz Graciano}
\author[1]{Al\'i Michel Angulo Mart\'inez}
\author[2*]{Dorilian Lopez-Mago}
\author[4]{Gustavo Castro-Olvera}
\author[4]{Martha Rosete-Aguilar}
\author[4]{Jes\'us Gardu\~no-Mej\'ia}
\author[3]{Roberto Ram\'irez Alarc\'on}
\author[1]{H\'ector Cruz Ram\'irez}
\author[1*]{Alfred B. U'Ren}

\affil[1]{Instituto de Ciencias Nucleares, Universidad Nacional Aut\'onoma de M\'exico, Apdo. Postal 70-543, Ciudad de M\'exico, M\'exico 04510}
\affil[2]{Tecnologico de Monterrey, Escuela de Ingenier\'ia y Ciencias, Ave. Eugenio Garza Sada 2501, Monterrey, N.L., M\'exico, 64849}
\affil[3]{Centro de Investigaciones en \'Optica A.C., Loma del Bosque 115, Colonia Lomas del Campestre, 37150 Le\'on Guanajuato, M\'exico}
\affil[4]{Instituto de Ciencias Aplicadas y Tecnolog\'ia, Universidad Nacional Aut\'onoma de M\'exico, Circuito Exterior, Cd. Universitaria, 04510, Ciudad de M\'exico, M\'exico}

\affil[*]{dlopezmago@tec.mx; alfred.uren@correo.nucleares.unam.mx}


\begin{abstract}
Optical-coherence tomography (OCT) is a technique that employs light in order to measure the internal structure of semi-transparent, e.g. biological, samples. It is based on the interference pattern of low-coherence light. Quantum-OCT (QOCT), instead, employs the correlation properties of entangled photon pairs, for example, generated by the process of spontaneous parametric downconversion (SPDC).  The usual QOCT scheme uses photon pairs characterised by a joint-spectral amplitude with strict spectral anti-correlations. It has been shown that, in contrast with its classical counterpart,  QOCT provides resolution enhancement and dispersion cancellation. In this paper, we revisit the theory of QOCT and extend the theoretical model so as to include photon pairs with arbitrary spectral correlations.  We present experimental results that complement the theory and explain the physical underpinnings appearing in the interference pattern.  In our experiment, we utilize a pump for the SPDC process ranging from continuous wave to pulsed in the femtosecond regime, and show that cross-correlation interference effects appearing for each pair of layers may be directly suppressed for a sufficiently large pump bandwidth.  Our results provide insights and strategies that could guide practical implementations of QOCT.
\end{abstract}

\begin{document}

\flushbottom
\maketitle
\thispagestyle{empty}

\section*{Introduction}

Optical coherence tomography (OCT) was introduced in 1991 by the group of Fujimoto\cite{Huang1991}. It is a non-invasive optical imaging technique that allows the acquisition of axial and cross-sectional images of semi-transparent objects, including biological samples. It uses low-coherence light or ultrashort laser pulses to measure the time of flight of light reflected from the internal structure of the sample. The time of flight information is reconstructed from the interference between the reflections and a reference arm. The usual configuration employed is a Michelson interferometer; the sample is located in one interferometer arm while the second arm is used as a reference. Reflections from the sample are combined with reflections from the reference mirror, and their interference is measured as a function of the position of the reference mirror. The axial resolution is mainly determined by the coherence of the probe light; for high-resolution images,  it is desirable to use light sources with a short coherence time. OCT has found several applications,  predominantly in the field of ophthalmology\cite{DrexlerandFujimoto2015}.

Before the advent of OCT, quantum interference experiments, perhaps best exemplified by the seminal work of Hong, Ou and Mandel (HOM),\cite{Hong1987} had gained prominence.  A HOM interferometer takes as input signal and idler photon pairs, produced for example, by the process of spontaneous parametric downconversion (SPDC) in a nonlinear crystal. The signal and idler photons are directed to the two input ports of a beamsplitter, and single-photon detectors monitor the rate of coincidence counts at the two output ports as a function of the delay between the two inputs.  
The resulting coincidence interferogram shows a dip, the so called HOM dip, centred at the zero input delay between the two photons.  The width of this interferogram determines the temporal width of the biphoton wavepacket. In comparison with a single-photon  wavepacket, the biphoton shows a narrower width by a factor of $1/2$, which implies an improvement in the resolution by a factor of two\cite{Abouraddy2002}. 

The theoretical work of Abouraddy \emph{et al.}\cite{Abouraddy2002} introduced the idea of exploiting the HOM experiment in order to realize a quantum version of OCT (referred to as QOCT). They experimentally demonstrated the idea a year later \cite{Nasr2003}, showing that QOCT presents some advantages over its classical counterpart, such as resolution enhancement and  dispersion cancellation \cite{Nasr2004}.  In QOCT, the HOM scheme is modified by placing the reflective sample in one of the arms of the interferometer, while the second arm, used as reference, is left intact. The mirror in the reference arm is translated and the coincidence interferogram is measured as a function of the resulting path-length difference. The interferogram shows the fourth-order interference of the photon pairs, which contains a HOM dip for each internal reflection in the sample. In addition, it has been shown that cross-correlation effects appear for each pair of sample layers \cite{Abouraddy2002,Nasr2003}.  

Several QOCT variations have been demonstrated\cite{Teich2011}. A polarization sensitive QOCT (PS-QOCT) was  demonstrated by Booth \textit{et al.} \cite{Booth2004, Booth2011}. Polarization-entangled photons from SPDC type II were used for axial sectioning and measurement of anisotropic properties. The technique measures the change in the state of polarization of the photon reflected from each sample layer. The change in the state of polarization arises from the birefringence in the sample material. Hence, PS-QOCT provides optical sectioning and information that could be used to describe the Jones matrix of each internal section of the sample.

Quantum-mimic OCT uses classical light with nonlinear effects to mimic the advantages of QOCT, such as dispersion cancellation and resolution enhancement\cite{Erkmen2006, Kaltenbaek2008,Kaltenbaek2009,Lavoie2009}. The use of classical light brings, as an important benefit, higher acquisition rates. It was suggested by these experiments that the advantages of QOCT are not due to the non-classical nature of the biphoton state, but to its phase-insensitive fourth-order interferogram. At the core of these quantum-mimic OCT schemes is phase conjugation, which is realized using an optical parametric amplifier. The reflected light from the sample passes through the amplifier and probes the sample again. This double-pass configuration provides the phase conjugation required to achieve dispersion cancellation \cite{Erkmen2006}. 

An implementation of QOCT including actual testing of a biological sample (onion skin tissue) was carried out by Nasr \emph{et al.} \cite{Nasr2009}. The sample was coated with spherical gold nanoparticles to increase its reflectance and reduce acquisition times.  Reconstruction of a three-dimensional image after collecting several interferograms was reported.  Later, quantum-mimic OCT also showed its potential for biological imaging \cite{Mazurek2013}.

The particular features exhibited by the HOM interferogram used for QOCT are dependent on the correlation properties inherent in the joint spectrum of the photon pairs. Previous QOCT schemes have been implemented using frequency entangled photons with strict frequency anti-correlations, derived from a continuous wave (CW)  pump in the SPDC process.  While it is known that the HOM dip characteristics, including width and visibility, are defined in part by the pump spectrum (which in turn determines the SPDC joint spectrum) \cite{Mazzotta2016}, to the best of our knowledge a full analysis of QOCT as a function of the shape of the joint spectrum has not been reported. The main objective of the present work is to analyse the interference effects that appear in a QOCT interferogram for different types of spectral correlations in the signal and idler photon pairs\cite{Torres2005,Vicent2010}.

We have developed a fully-automated and fibre-coupled QOCT experiment (the experiment is fibre-based except for a temporal delay in free space, and the interaction of one of the photons with the sample, evidently also in free space).   We are able to shape the photon-pair spectral correlations by controlling the pump bandwidth. We can directly ascertain the shaped spectral correlations in the SPDC photon pairs by measuring the joint spectrum\cite{Zielnicki2018}. We have performed a theoretical and experimental study of how the joint spectrum determines the HOM interferogram used for QOCT. We have found the conditions under which the cross-correlation effects from each pair of sample layers, obtained in QOCT when using entangled photons, can be eliminated. We also demonstrate a complementary technique used for measuring \emph{both} the thickness and refractive index of a two-layer sample. Furthermore, in our experiments, the photon pairs are centred at $1550$ nm, which is to the best of our knowledge, the first experimental demonstration of QOCT in the telecommunications band.

\section*{Theory}

\subsection*{QOCT with frequency-correlated photons}  

\begin{figure}[htbp]
\centering
\includegraphics[width=13cm]{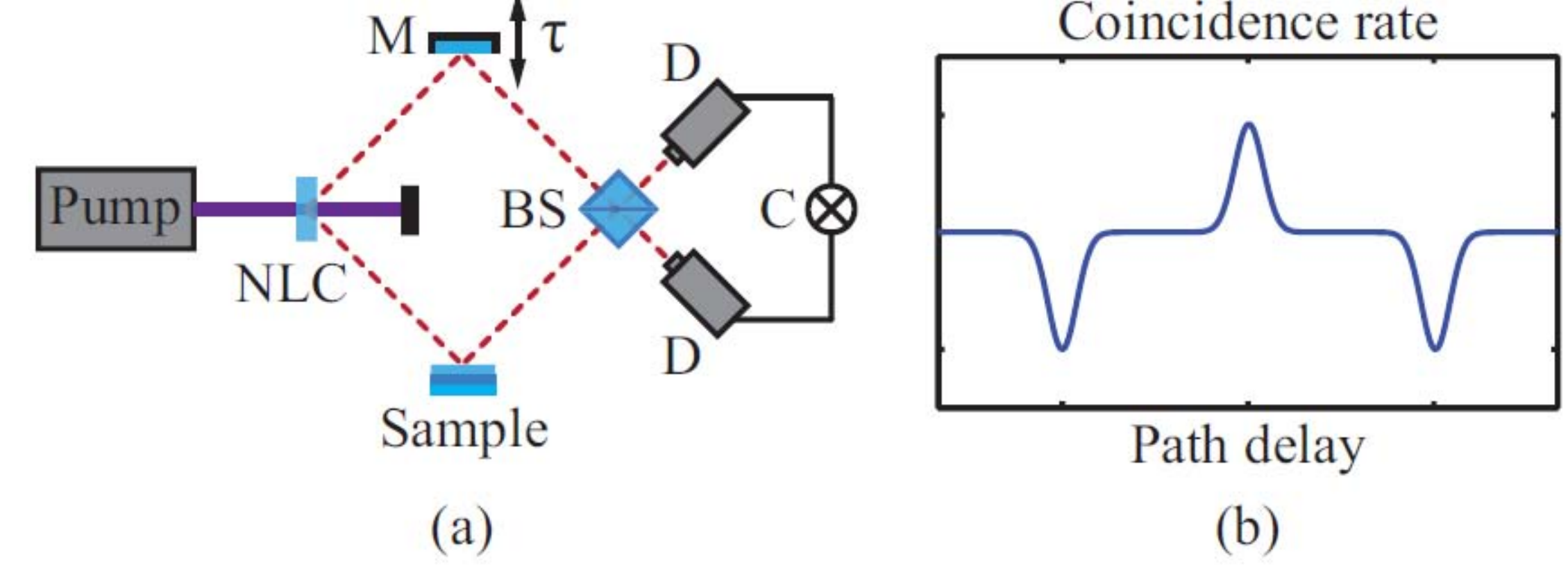}
\caption{(Color online) (a) Schematic diagram of QOCT using SPDC photon pairs. Signal photons reflected from the sample interfere at the beam splitter (BS) with idler photons carrying a temporal delay $\tau$. (b) Example of a QOCT interferogram for a two-layer sample. The separation of the dips is proportional to the sample optical thickness. The central peak is due to the cross-correlation between the reflection of the probe photon by both layers.}
\label{fig:schematic}
\end{figure}

Figure \ref{fig:schematic}(a) shows a schematic of the QOCT apparatus, employing a HOM interferometer. In the original experiment realized by Abouraddy \emph{et al.}\cite{Abouraddy2002} a CW laser pumps a nonlinear crystal to generate broadband entangled photon pairs via type I SPDC.  The signal photon probes the sample and interferes at a beam splitter with the idler photon; the two photons enter the beamsplitter through separate input ports with a temporal delay $\tau$. Single-photon detectors in combination with coincidence electronics monitor coincidence detection events at the two outputs of the beam splitter as a function of $\tau$, thus yielding the QOCT interferogram $C(\tau)$. Figure \ref{fig:schematic}(b) shows a schematic example of a coincidence interferogram for a two-layer sample. This interferogram features two dips and a central structure, which as discussed below can be either a peak or a dip.  While the two dips are produced by destructive quantum interference of the reflections from the two consecutive interfaces, the central structure arises from interference between both reflections. 

The sample to be analysed with the QOCT device is characterised by the function $H(\omega)$, henceforth referred to as the sample reflectivity function (SRF) expressed as 
\begin{equation}
H(\omega) = \int \mathrm{d}z\, r(z,\omega) \exp (i\phi(z,\omega) ).
\end{equation}

In cases where the sample is composed of $N$ discrete layers, the SRF  function $H(\omega)$  is obtained by adding the contributions from all layers, as follows
\begin{equation}
H(\omega)= \sum\limits_{j=0}^{N-1} r_{j} e^{i\omega T_{j}},
\end{equation}
\noindent where $N$ is the number of layers (here a layer is understood as an interface between two refractive indices), $T_{j}$ is the time traveled to the $j$th layer and back,  while $r_{j}$ is the corresponding reflectivity.

\subsection*{QOCT with spectrally engineered photon pairs} 

Let us consider the case of a  pulsed pump; the monochromatic pump case can then be obtained as a special case in the appropriate limit.   In this situation, the pump bandwidth is non-vanishing and  the two-photon state can be expressed as
\begin{equation}
\ket{\psi} = \ket{0}_1\ket{0}_2+ \eta \int \mathrm{d}\omega_1 \int \mathrm{d}\omega_2 f(\omega_1,\omega_2) \ket{\omega_{1}}\ket{\omega_{2}},
\end{equation}
\noindent with a two-dimensional joint spectral amplitude $f(\omega_{1},\omega_{2})$ where $|\eta|^2$ is proportional to the photon-pair generation rate.  Note that the (normalised) joint spectral intensity (JSI) or joint spectrum $|f(\omega_{1},\omega_{2})|^{2}$ can be interpreted as the probability of detecting a signal photon with frequency $\omega_{1}$ and its idler photon with frequency $\omega_{2}$. 
Refer to the schematic of the QOCT experiment shown in Fig. \ref{fig:schematic}(a). The coincidence interferogram is obtained by the expression (details of the derivation in the Methods section)
\begin{equation}\label{Eq:coinc_interf}
C(\tau)  = \frac{1}{4} \int \int \mathrm{d}\omega_{1}  \mathrm{d}\omega_{2} \left| f(\omega_{2},\omega_{1})H(\omega_{1})e^{i(\omega_{2}-\omega_{1})\tau} - f(\omega_{1},\omega_{2})H(\omega_{2})  \right|^{2}. 
\end{equation}

By expanding equation (\ref{Eq:coinc_interf}), it is found that $C(\tau)$ has the general form 
\begin{equation} \label{HOM}
C(\tau)=A+B-G(\tau)-G^{\ast}(\tau),
\end{equation}
\noindent where
\begin{equation}\label{Eq:AandB}
A  = \frac{1}{4} \int \mathrm{d}\omega_{1} \int \mathrm{d}\omega_{2} |f(\omega_{2},\omega_{1})H(\omega_{1})|^{2}, \quad
B  = \frac{1}{4} \int \mathrm{d}\omega_{1} \int \mathrm{d}\omega_{2} |f(\omega_{1},\omega_{2})H(\omega_{2})|^{2},
\end{equation}
are recognized as self-interference terms (equivalent to $\Lambda_{0}$ in Abouraddy \emph{et. al.}\cite{Abouraddy2002}), independent of $\tau$, and
\begin{equation}\label{Eq:crossinterf}
G(\tau) = \frac{1}{4} \int \mathrm{d}\omega_{1} \int \mathrm{d}\omega_{2} f(\omega_{2},\omega_{1})f^{\ast}(\omega_{1},\omega_{2})H(\omega_{1})H^{\ast}(\omega_{2})e^{i(\omega_{2}-\omega_{1})\tau},
\end{equation}
is identified as the cross-interference term (equivalent to $\Lambda(\tau)$\cite{Abouraddy2002}), which contains the information related to the internal structure of the sample. In order to understand the features of $G(\tau)$, we find it convenient to work with the joint temporal amplitude $\tilde{f}(t_{1},t_{2})$, which is related to $f(\omega_{1},\omega_{2})$ through a Fourier transform as follows
\begin{equation}\label{Eq:JTA}
f(\omega_{1},\omega_{2}) = \int \mathrm{d}t_{1} \int \mathrm{d}t_{2} \tilde{f}(t_{1},t_{2}) e^{i\omega_{1}t_{1}}e^{i\omega_{2}t_{2}}.
\end{equation}
Using equations (\ref{Eq:JTA}) and (\ref{Eq:crossinterf}) and substituting the discretized version of $H(\omega)$, we find that
\begin{equation}\label{Eq:Gtau}
G(\tau) = \frac{1}{4} \sum_{j=0}^{N-1}  \sum_{l=0}^{N-1} r_{j} r^{\ast}_{l} \int \int \mathrm{d}t_{1}  \mathrm{d}t_{2} \tilde{f}(t_{2},t_{1}) \tilde{f}^{\ast}(t_{1}-\tau + T_{j},t_{2}+\tau-T_{l}).
\end{equation}

Note that the interferogram is given by equation (\ref{HOM}) in general form, i.e. without assuming a particular form for the joint spectrum. We now employ a simplified model for the joint spectrum in order to gain some understanding of the features observed in the interferogram, as determined by the type of spectral correlations present in $f(\omega_{1},\omega_{2})$. Let us consider a joint spectral amplitude of the form
\begin{equation} \label{Eq:JoinSpectrumModel}
f(\omega_{1},\omega_{2})=\frac{2}{ \sqrt{\pi \Omega_{a} \Omega_{d} }}\exp\left[-\left(\frac{\omega_{1}-\omega_{2}}{\Omega_{a}}\right)^{2}\right] \exp\left[-\left(\frac{\omega_{1}+\omega_{2}-2\omega_{0}}{\Omega_d}\right)^{2}\right].
\end{equation}

\begin{figure}[htbp]
\centering
\includegraphics[width=14cm]{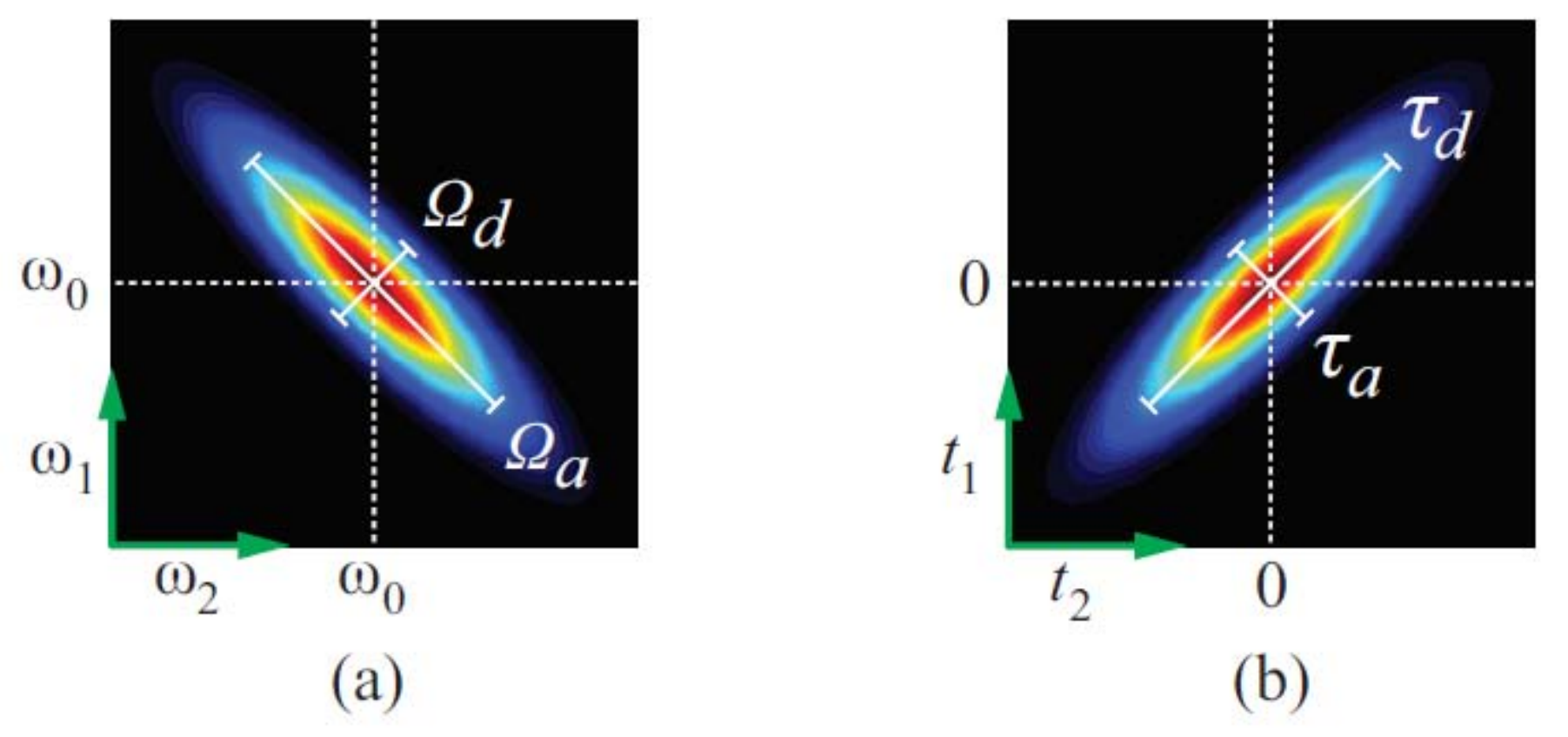}
\caption{(Color online). (a) Qualitative picture for the joint spectrum $|f(\omega_{1},\omega_{2})|^{2}$ given in equation (\ref{Eq:JoinSpectrumModel}). (b) Qualitative picture of the joint temporal function $|\tilde{f}(t_{1},t_{2})|^{2}$ given in equation (\ref{Eq:JoinTempAmp_withtaus}).}
\label{fig:JSImodel}
\end{figure}

A schematic of the joint spectrum resulting from this simplified model is presented in Fig. \ref{fig:JSImodel}(a). It is given by the product of two Gaussian functions, in general with dissimilar $1/e$ (intensity) full width parameters $\Omega_a$ and $\Omega_d$  along the anti-diagonal and diagonal  directions, respectively, in $\{\omega_1,\omega_2\}$ space, yielding an ellipse with its axes parallel to the diagonal and anti-diagonal. The normalisation factor $2/\sqrt{\Omega_{a} \Omega_{d} \pi}$ ensures that $\int \int |f(\omega_{1},\omega_{2})|^{2} \mathrm{d}\omega_{1} \mathrm{d}\omega_{2} = 1$. The joint spectrum is centred at $(\omega_{1},\omega_{2})=(\omega_{0},\omega_{0})$, with $\omega_{0}=\omega_{p}/2$, where $\omega_{p}$ is the central pump frequency. 

The corresponding joint temporal amplitude, in the absence of group velocity dispersion effects, is then given by
\begin{equation} \label{Eq:JoinTempAmp_withtaus}
\tilde{f}(t_{1},t_{2}) = \frac{2}{\sqrt{\pi\tau_{a}\tau_{d}}}\exp\left[i \omega_{0} (t_{1}+t_{2})\right] \exp\left[-\left(\frac{t_{1}-t_{2}}{\tau_{a}}\right)^{2}\right]\exp\left[-\left(\frac{t_{1}+t_{2}}{\tau_{d}}\right)^{2}\right],
\end{equation}
\noindent characterised by $1/e$ (intensity) anti-diagonal and diagonal  temporal full width parameters $\tau_a$ and $\tau_d$, respectively.   We refer to $\tau_a$ and $\tau_d$ as the anti-diagonal and diagonal entanglement times and they are given in terms of the spectral widths $\Omega_a$ and $\Omega_d$ as follows
\begin{equation}
\tau_a=\frac{4}{\Omega_a}, \quad \tau_d=\frac{4}{\Omega_d}.
\end{equation}

These two time parameters define, together, the time-frequency entanglement properties of our source.  The joint temporal intensity corresponding to the joint spectral amplitude in Fig. \ref{fig:JSImodel}(a) is shown schematically in Fig. \ref{fig:JSImodel}(b).   Note the inverse Fourier relationship  between the diagonal/anti-diagonal widths  in the spectral and temporal domains.  

In order to gain further insight, let us consider as an example a two-layer sample, leading to an SRF of the form 

\begin{equation}
H(\omega)=r_{0}e^{i \omega T_{0}} + r_{1} e^{i \omega T_{1}}.
\end{equation}

\noindent where $r_0$ and $r_1$ represent the reflectivities of the two layers and  $(T_1-T_0)c$ represents the optical path difference between the two layers. Using equations (\ref{Eq:AandB}) and (\ref{Eq:Gtau}) together with equations (\ref{Eq:JoinSpectrumModel}) and (\ref{Eq:JoinTempAmp_withtaus}) we find that the QOCT interferogram $C(\tau)$, normalised so that $C(\tau)$=1 for $\tau \rightarrow \pm \infty$, can be expressed as

\begin{equation}\label{Eq:Cfortwolayers}
\frac{C(\tau)}{A+B} = 1-V_{0}\exp\left[-2\left(\frac{\tau}{\tau_{a}}\right)^{2}\right]-V_{\mathrm{mid}}\exp\left[-2\left(\frac{\tau-T/2}{\tau_{a}}\right)^{2}\right]-V_{1}\exp\left[-2\left(\frac{\tau-T}{\tau_{a}}\right)^{2}\right],
\end{equation}

\noindent where $T=T_{1}-T_{0}$ is the propagation time between the two layers and
\begin{equation}\label{Eq:visibilities}
V_{0}=\frac{R_{0}}{2(A+B)}, \quad V_{1}=\frac{R_{1}}{2(A+B)}, \quad V_{\mathrm{mid}} =\frac{\sqrt{R_{0}R_{1}}}{A+B} \exp[-\frac{1}{2}\left(\frac{T}{\tau_{d}}\right)^{2}]\cos(\omega_{0}T),
\end{equation}
\noindent with
\begin{equation}\label{Eq:AplusB}
A+B = \frac{1}{2}(R_{0}+R_{1})+\sqrt{ R_{0}R_{1}}\exp \left[-\left(\frac{T}{\tau_{a}}\right)^{2}\right] \exp\left[-\left( \frac{T}{\tau_{d}}\right)^{2}\right]\cos(\omega_{0}T) \approx \frac{1}{2}(R_{0}+R_{1}),
\end{equation}
\noindent in terms of the reflectivities of the first and second interfaces  $R_{0} = |r_{0}|^{2}$ and $R_{1} = |r_{1}|^{2}$. Note that the second term in equation (\ref{Eq:AplusB}) can be neglected when either $T \gg \tau_a$ or $T \gg \tau_d$.

Equation (\ref{Eq:Cfortwolayers}) shows three elements:  two dips separated by a time $T=2 n L /c$ (with $L$ the sample thickness and $n$ its refractive index), with visibilities $V_{0}$ and $V_{1}$, as well as an intermediate structure located at $\tau=T/2$, \emph{i.e.} midway between the two dips,  with amplitude $V_{\mathrm{mid}}$.  Note that each of the two dips is associated with one of the two interfaces in the sample, while the intermediate structure is associated with both interfaces.
We can see from the expression for  $V_{\mathrm{mid}}$, equation (\ref{Eq:visibilities}), that this structure can be either a dip or a peak, depending on the sign of $\mbox{cos}(\omega_0 T)$, with an amplitude determined by the various source and sample parameters.

Let us turn our discussion to how the entanglement time parameters $\tau_a$ and $\tau_d$ are determined in various experimental situations of interest.  The value of $\tau_d$ is largely determined by the pump temporal properties.   On the one hand, if the pump is monochromatic the diagonal spectral width $\Omega_d$ vanishes and $\tau_d \rightarrow \infty$: thus, evidently in this case $\tau_d \gg T$.  On the other hand, if the pump is in the form of an ultrashort (e.g. femtosecond) pulse train, and assuming that the diagonal spectral width is limited by the pump spectral amplitude rather than the phasematching function, we can express $\tau_d$ in terms of the pump pulse duration $\tau_p$ as $\tau_d=\sqrt{2} \tau_p$.   For $L=1$ mm and $\tau_p =100$ fs, we obtain $\tau_d/T \approx 0.014$.  Thus, interestingly, while $\tau_d \gg T$ for the CW case, this relationship can be inverted to $\tau_d \ll T$ for the ultrashort pulsed pump case.   

While the anti-diagonal spectral width $\Omega_a$ is of course influenced by the crystal phasematching properties as well as pump properties, in our case the resulting anti-diagonal spectral width is limited by the bandpass filter used (which in our case is centred at $1550$ nm with a  $7.8$ nm $1/e$ bandwidth). From an inspection of equation (\ref{Eq:Cfortwolayers}), it becomes evident that in an experimental measurement of the QOCT interferogram, $\tau_a$ can  be directly obtained from the HOM dip width; $\tau_a$ essentially corresponds to the dip width.   Note that in order for the QOCT measurement to yield meaningful information about the sample, the dip separation, which corresponds to $T$, must be considerably larger than $\tau_a$.   Indeed, if $T \lesssim \tau_a$, the two dips cannot be resolved and we are unable to extract useful information about the sample. 

\begin{figure}[htbp]
\centering
\includegraphics[width=5cm]{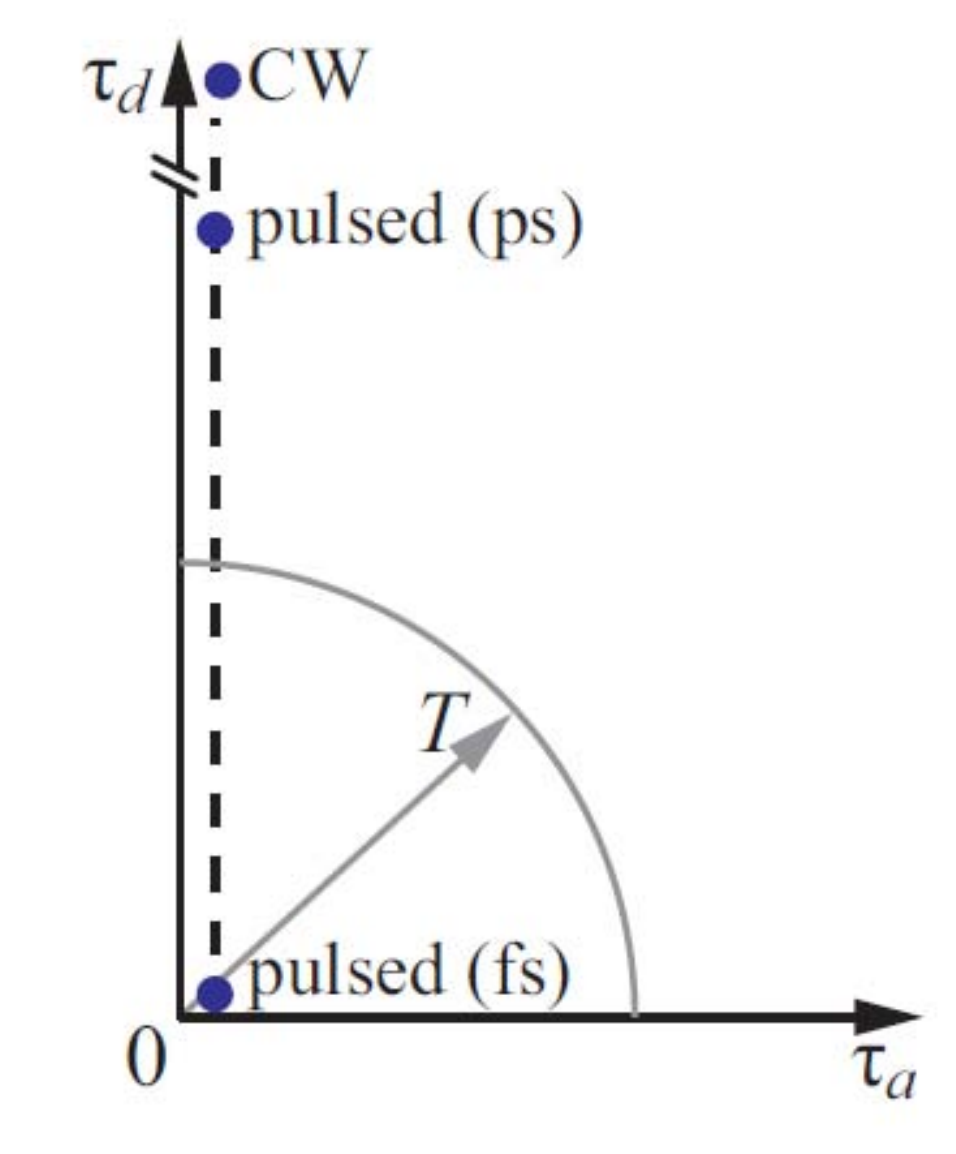}
\caption{(Color online). Schematic diagram representing the time scales used in the experiment.}
\label{fig:timesdiagram}
\end{figure}

In order to understand the interplay between the two-photon state (parameters $\tau_a$ and $\tau_d$) and the sample thickness (parameter $T$) in defining the resulting QOCT interferogram, it is useful to visualise the parameter space $\{\tau_a,\tau_d\}$.  Note that in the Gaussian model being employed for the joint spectral amplitude (see equation (\ref{Eq:JoinTempAmp_withtaus})), the state is fully determined by the coordinates in this space.   An inspection of the expressions for the dip visibilities and the intermediate structure amplitude, equations  (\ref{Eq:visibilities}) and (\ref{Eq:AplusB}), reveals that $\tau_a$ and $\tau_d$ appear only  as quotients $\tau_a/T$ and    $\tau_d/T$.  Therefore, in Fig. \ref{fig:timesdiagram} we indicate the sample thickness as a circle of radius $T$; for this illustration we have assumed a value of $T$ which corresponds as in our experiment to be described below, in a double-pass arrangement, to a $1$ mm thickness of soda lime glass (with a resulting value of $T\approx10$ ps). We have indicated with a vertical dashed line the value of $\tau_a$, which is essentially the same irrespective of the pump bandwidth.

In Table \ref{Tab:parameters} we show for three different sources (CW pump, picosecond-duration pump with $\tau_p\approx10$ ps and a femtosecond-duration pump with $\tau_p\approx100$ fs) the resulting values of $\tau_d$ and $\tau_a$, which correspond to the sources used in our experimental measurements to be discussed below.  We have placed markers in Fig. \ref{fig:timesdiagram} for these three sources.

With the values of $\tau_a$ in Table \ref{Tab:parameters}, it can be verified that $\exp[-(T/\tau_{a})^{2}]$ ends up being a small number so that the cosine-dependent term in equation (\ref{Eq:AplusB}) can be ignored, yielding $A+B=(R_0+R_1)/2$.  This in turn leads to the following expressions for the amplitudes of the two dips:
\begin{equation}\label{E:CWvis}
V_0=\frac{R_0}{R_0+R_1}, \quad V_1=\frac{R_1}{R_0+R_1}.
\end{equation}
 
\begin{table}
\caption[Table caption text]{Source parameters}
\centering
\begin{tabular}{ c | c | c | c  }
  \hline \hline
   source configuration & $\Delta \lambda_{p} (\mbox{nm})$ & $\tau_{d} (\mbox{ps})$ &  $\tau_{a} (\mbox{ps})$ \\
  \hline \hline
  CW pump & $0.12$ & $(\tau_d \to \infty)$ & $0.499 \pm 0.028$   \\
  pulsed pump (ps) & $0.30$ & $17.006$ & $0.440 \pm 0.014$ \\
  pulsed pump (fs) & $10.8$ & $0.472$ & $0.432 \pm 0.026$ \\
  \hline  
\end{tabular} 
\label{Tab:parameters}
\end{table}
 
Note that in the monochromatic pump limit, $\exp[-(T/\tau_{d})^{2}]\rightarrow 0$ is fulfilled and the expressions above (see equation (\ref{E:CWvis})) for the dip visibilities are therefore valid.  As for the amplitude of the intermediate structure, it clearly depends on the quotient $T/\tau_d$.  In the monochromatic case for which $\tau_d \rightarrow \infty$, we obtain
\begin{equation}\label{E:CWintstru}
V_{\mathrm{mid}}= \frac{2\sqrt{R_0 R_1}}{R_0+R_1}\mbox{cos}(\omega_0 T),
\end{equation}
\noindent while in the ultrashort pulsed pump case, for which $T \gg \tau_d$, we obtain $V_{\mathrm{mid}}=0$. The criterion for suppression of the intermediate structure can be expressed as a threshold for the diagonal spectral width, as follows, in terms of the sample width $L$:
\begin{equation} \label{E:condition}
\Omega_d \gtrsim \frac{2 c}{n L}.
\end{equation}
Thus, as $\Omega_{d}$ is allowed to increase from a value of $0$, for example replacing a CW pump laser with a pulsed laser,  the visibility of the intermediate structure gradually decreases until it is altogether suppressed as already discussed in cases where the above inequality is fulfilled.  Note that a factorable two-photon state  obtained through group velocity matching\cite{Vicent2010} will typically fulfill the inequality in equation (\ref{E:condition}).

Note that in the specific case of equal reflectivities, i.e. $R_0=R_1$, with a monochromatic pump we obtain $V_0=V_1=1/2$ and $V_{\mathrm{mid}}=\mbox{cos}(\omega_0 T)$.  Note also that as the number of layers increases, the visibility of the HOM dip associated with each layer will be reduced as $1/N$, where $N$ is the number of layers.  Therefore, while the ideal visibility is reduced from $1$, in the standard HOM interferometer, to $0.5$, in the two-layer QOCT with equal reflectivities, the intermediate structure with a CW pump constitutes either a peak or a dip, as governed by $\mbox{cos}(\omega_0 T)$, with an amplitude which does not decrease with the number of layers.   For a fixed sample width $T$ we are able to control the amplitude, and the dip/peak character, of the intermediate structure through the pump frequency $\omega_{p}=2\omega_{0}$.

\subsection*{QOCT with photon pairs from SPDC with a continuous-wave pump}  

In general, the joint spectral amplitude function $f(\omega_s,\omega_i)$ can be expressed as $\phi(\omega_s,\omega_i) \alpha(\omega_s+\omega_i)$ where $\phi(\omega_s,\omega_i)$ represents the phasematching function and  $\alpha(\omega_s+\omega_i)$ is the pump envelope function\cite{grice97}. In the monochromatic pump limit, $\alpha(\omega_s+\omega_i)=\delta(\omega_s+\omega_i-\omega_p)$, where $\omega_p$ is the pump frequency, so that the two-photon component of the state is now described, in terms of a single integral,  as follows:
\begin{equation}\label{E:state_m}
\ket{\psi_m(\omega_p)} =  \int \mathrm{d} \Omega \phi\left(\frac{\omega_p}{2}+\Omega,\frac{\omega_p}{2}-\Omega\right)|\frac{\omega_p}{2}+\Omega \rangle|\frac{\omega_p}{2}-\Omega \rangle.
\end{equation}

For such a state, with SPDC photon pairs obtained with a monochromatic pump, the QOCT interferogram can be written as
\begin{equation}
C_m(\tau; \omega_p )=\int d \Omega \left| \phi\left(\frac{\omega_p}{2}+\Omega,\frac{\omega_p}{2}-\Omega\right)H\left(\frac{\omega_p}{2}-\Omega\right)-\phi\left(\frac{\omega_p}{2}-\Omega,\frac{\omega_p}{2}+\Omega\right)H\left(\frac{\omega_p}{2}+\Omega\right)e^{i 2 \Omega \tau}  \right|^2.
\end{equation}

Note that for the specific case of the Gaussian model for the two-photon state used above (see equation (\ref{Eq:JoinSpectrumModel})), in the monochromatic pump limit, the resulting interferogram has already been obtained in the limit $\Omega_d \rightarrow 0$ (or equivalently $\tau_d \rightarrow \infty$).  It is given by equation (\ref{Eq:Cfortwolayers}) with the visibilities of the two HOM dips and amplitude of the intermediate structure given by equations (\ref{E:CWvis}) and (\ref{E:CWintstru}), respectively.

In practice, the CW laser used as pump has a non-vanishing bandwidth; thus, in the case of a realistic CW pump laser with a spectral linewidth function $\alpha_{cw}(\omega)$, the resulting SPDC two-photon state can be expressed as a statistical mixture of the  pure states associated with each pump spectral component.  The state is then described through a density operator as follows:
\begin{equation}
\hat{\rho}=\int d \omega_p \alpha_{cw}(\omega_p) \ket{\psi_m(\omega_p)} \bra{\psi_m(\omega_p)}
\end{equation}
\noindent where $\alpha_{cw}(\omega)$ is normalised so that the integral over all $\omega$ of $|\alpha_{cw}(\omega)|^2$ yields unity. It can be shown that in this case, the QOCT interferogram is obtained through the weighted average, with $\alpha_{cw}(\omega)$ as weighting factor, of the interferogram corresponding to each pump spectral component, as follows:
\begin{equation}\label{E:interfer_CW}
C_{cw}(\tau)=\int d \omega_p \alpha_{cw}(\omega_p) C_m(\tau; \omega_p ).
\end{equation}

Because the amplitude of the intermediate structure is proportional to $\mbox{cos}(\omega_0 T)$, the averaging over the pump spectral components in equation (\ref{E:interfer_CW}) will tend to suppress the intermediate structure. Note that this is a similar effect to that obtained for a pulsed pump; however, the typical CW laser linewidths (such as the one used in our experiments) are insufficient to fully suppress the intermediate structure.

\section*{Results}

\subsection*{Experimental setup} 

\begin{figure}[ht!]
\centering
\includegraphics[width=15 cm]{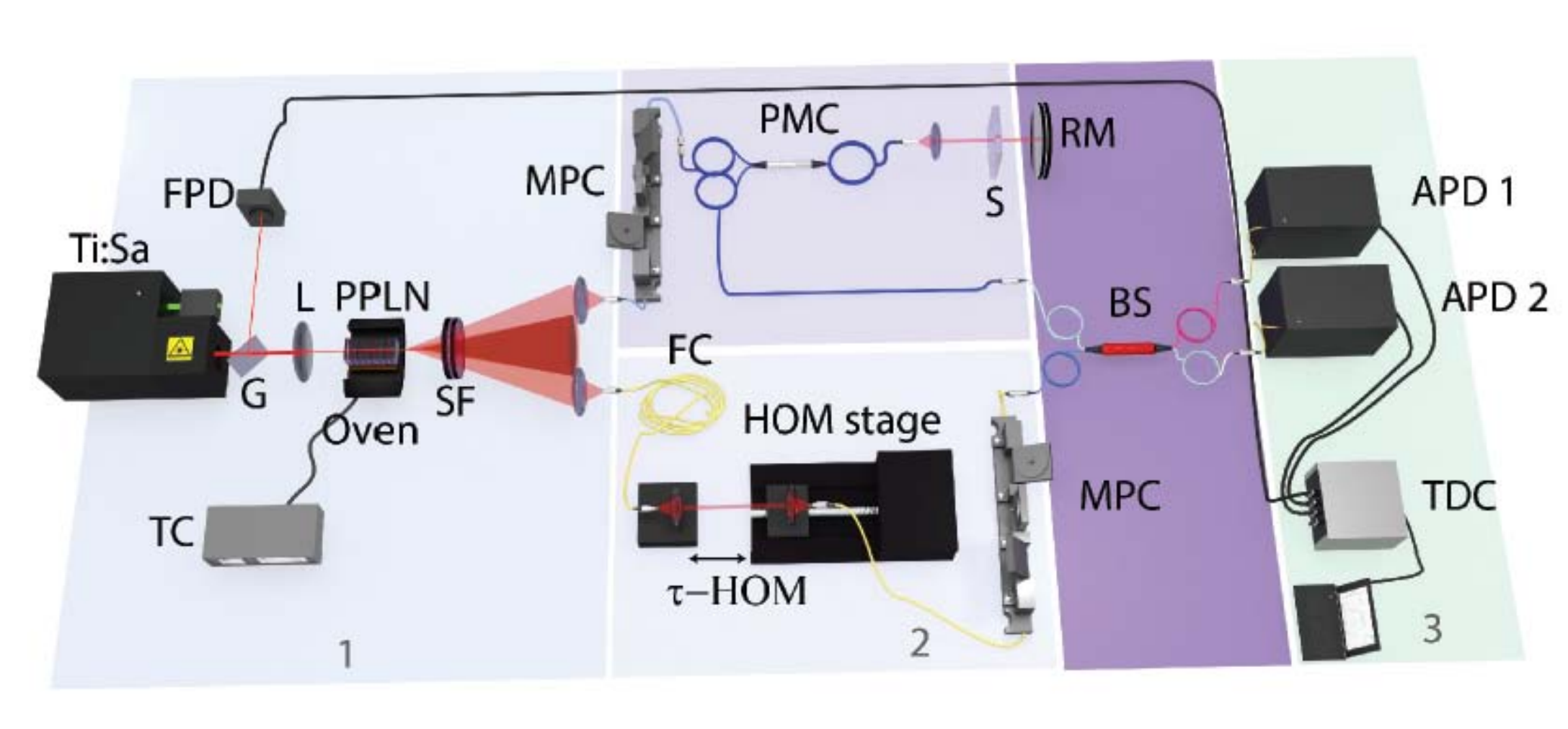}
\caption{(Color online) Experimental setup designed to measure the QOCT interferograms. Part 1: TC: temperature controller, L: plano-convex spherical lens, G: glass beam sampler, FPD: fast photodiode, PPLN: periodically poled lithium niobate crystal, SF: set of bandpass and longpass filters. Part 2: MPC: manual fibre polarization controller, PMC: polarization maintaining optical circulator, FC: compensating fibre, S: sample, RM: reference mirror, BS: beam splitter. Part 3: TDC: time to digital converter, APD: avalanche photodetectors. Details of the elements are explained in the Methods section.}
\label{fig:setup}
\end{figure}

In order to validate the theoretical description presented above, we have carried out an experiment; our setup is shown in Fig. \ref{fig:setup}. We have used three configurations, as shown in Table \ref{Tab:parameters}, designed to range from continuous wave to the ultrashort, femtosecond pulsed pump case.  These three configurations are based on two lasers, both centred at $775$ nm: a femtosecond Ti:Sapphire oscillator, as well as a picosecond Ti:Sapphire oscillator which can be operated either in pulsed or CW mode.

We use a PPLN crystal of $10$ mm length, with poling period of $19.1$ $\mu$m, and operated at a temperature of $90^{\circ}$ C. This crystal is pumped according to each of the three configurations described in the previous paragraph.  Our SPDC source produces frequency-degenerate, non-collinear photon pairs (propagating at $\pm 1.2^\circ$ with respect to the pump mode in free space), centred at a wavelength of $1550$ nm. We couple each of the photons of a given pair into a single mode fibre, after being transmitted through a bandpass filter centred at $1550$ nm with $7.8$ nm 1/e full width.  We manipulate each of the photons in a particular manner before interfering them at a fibre 50:50 beamsplitter.  The signal photon is injected into one of the ports of a fibre circulator, it exits the circulator into free space through a second port, passes through a collimating lens, and reaches the sample. If the single photon is reflected from the sample, it is injected back into the same port of the circulator, and finally emerges into a fibre connected to the third port. The idler photon is coupled out from the fibre into free space and coupled back into another fibre mounted on a computer controlled translation stage, so that we are able to introduce an arbitrary temporal delay $\tau$ between the two photons.    After interfering at the beamsplitter (the two arms are of course balanced so that they have equal length), the two photons emerging from the two output ports of the beamsplitter are directed to free-running fibre-coupled InGaAs avalanche photodiodes.  We have used fibre polarization controllers in both arms so as to make sure that the polarizations reaching the beamsplitter are identical.  Coincidence counts are monitored with the help of a time to digital converter (time tagger) as a function of the time delay $\tau$.

\subsection*{Control and characterisation of the joint spectrum} 

A number of methods exist for controlling the joint spectrum of SPDC photon pairs\cite{Torres2011}. For a crystal with given properties, it can be achieved by controlling: the pump waist $w_{0p}$ and/or the diameter of the collection modes at the nonlinear crystal \cite{Vicent2010, Gerrits2011},  the pump bandwidth $\Omega_{p}$ \cite{Eckstein2011, Lutz2014}, or the pump angular dispersion\cite{Hendrych2007}. In our experiment, we control the joint spectrum by selecting the pump bandwidth (through our choice of pump configurations amongst three different possibilities), together with transmission of our photon pairs through a bandpass filter.

In order to characterise the joint spectrum we built a fibre spectrometer\cite{Lutz2014, Zielnicki2018}. In this device, we exploit group velocity dispersion in a long fibre so that each spectral component is  delayed by a frequency-dependent time; with appropriate calibration, time of arrival information can then be converted to a measured spectrum.  For this spectrometer (not shown in Fig. \ref{fig:setup})  we transmit each photon through a $5$ km length of a SMF28 fibre before being detected by an InGaAs APD.  For the pulsed-pump case, a small portion  of the train pulse from the laser is reflected and detected with a fast photodiode. Time tags from each of the two detectors $t_1$ and $t_2$ are subtracted from the corresponding time tag from the fast photodiode $t_0$. Following a calibration step, we translate time of arrival information,  specifically $t_1-t_0$ and $t_2-t_0$, to a frequency for each of the two photons. 2D histograms of times of arrival can thus be converted to measured joint spectral intensities. In the CW pump case, we no longer have an external time reference, and the time difference $t_1-t_2$ can be translated following a calibration step to the frequency detuning $\Omega$ (see equation (\ref{E:state_m})) or to the (marginal) signal frequency $\omega_s$.    In this case, a 1D histogram of time of arrival differences can be translated into the 1D JSI.  Note that the jitter of our APD's implies that spectral features with a width less than approximately $3$ nm along the diagonal and $5$ nm along the antidiagonal directions cannot be resolved.

We have prepared our two-photon state in three different configurations corresponding to our SPDC crystal pumped with: i) a CW pump, ii) a pulsed pump in the ps regime, and iii) a pulsed pump in the fs regime. The measured pump bandwidth used for these three configurations is shown in the first column of Table \ref{Tab:parameters}. We have shown the measured JSI  for each of these three cases  in the first column of Fig. \ref{fig:QOCTresults}. Note that, while the measured  JSI for the CW case is one-dimensional, it has been displayed for ease of comparison with the other two cases along the anti-diagonal in $\{\omega_s,\omega_i\}$ space (so that the width of the pixels along the opposite, i.e. diagonal, direction has no physical meaning).  In all three measurements our photon pairs are transmitted through  a  bandpass filter centred at $1550$ nm with $7.8$ nm $1/e$ full width, following the nonlinear crystal.  Note that the apparent  width, along the diagonal direction, of the JSI for the ps pump is considerably greater than the actual value (indicated in each case by a pair of diagonal parallel lines) because of the non-ideal resolution of the fibre spectrometer.

It is important to note that the type of spectral correlations which characterise our photon pairs can be controlled by the pump bandwidth.  While for the CW pump our source exhibits marked spectral anti-correlations, these become less marked for the ps pulsed case.  In contrast, note that for the fs pulsed pump case the resulting JSI is approximately factorable as can be appreciated from Fig. \ref{fig:QOCTresults}(c). 
\subsection*{QOCT measurements}

\begin{figure}[htbp]
\centering
\includegraphics[width=15cm]{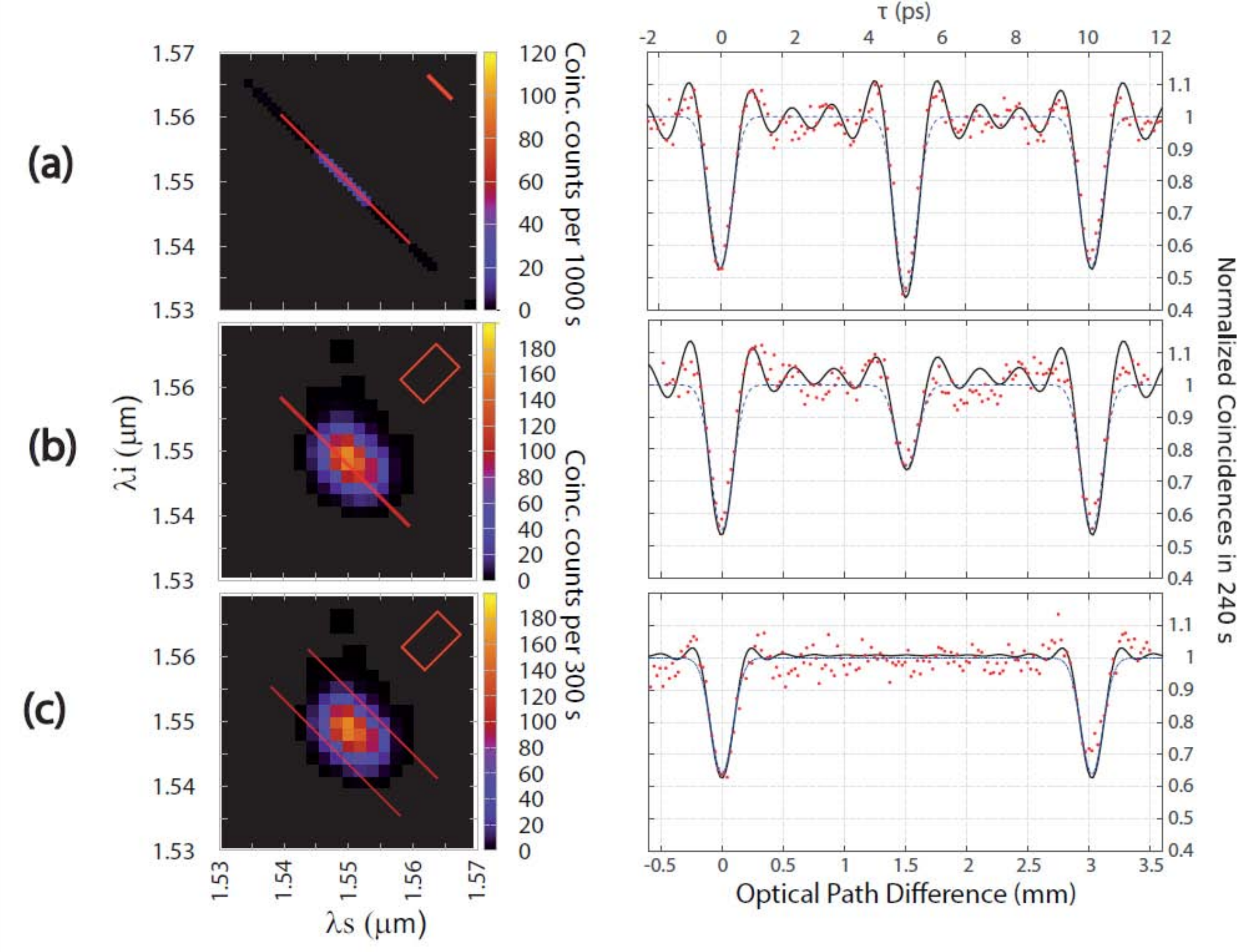}
\caption{(Color online) QOCT measurements of a two-layer sample for different JSI configurations, as follows:  (a) CW pump . (b) pulsed ps pump. (c) pulsed fs pump.}
\label{fig:QOCTresults}
\end{figure}

We have used as sample a microscope slide with a nominal thickness of $1$ mm and reflectance of about $4\%$ at each interface. An interferogram is recorded by monitoring the signal and idler coincidence detection rate as a function of the free-space delay. We use scanning steps of $2.5$ $\mu$m with acquisition times of $240$ s per delay setting.   The experimental QOCT interferograms for this sample are shown in the second column of Fig. \ref{fig:QOCTresults} for our three source configurations.  For each source configuration, we have shown experimental measurements (red dots), as well as two  theory curves: the interferogram derived from the Gaussian model (equations (\ref{Eq:JoinSpectrumModel})-(\ref{Eq:JoinTempAmp_withtaus})) (dashed blue line), and the interferogram derived from numerical integration of equations (\ref{Eq:AandB}) and (\ref{Eq:crossinterf}) with a full model\cite{Vicent2010}, i.e. including the presence of spectral filters (solid black line).  Note that while, unsurprisingly, the dips and intermediate structure exhibit a Gaussian shape in the case where our source is modeled through a Gaussian function (see equations (\ref{Eq:JoinSpectrumModel})), the fact that the spectral filter used has roughly a rectangular transmission function results in the sinc-type oscillations which surround each of the three structures.  

Table \ref{Tab:parameters}  shows the time entanglement parameters  $\tau_d$ and $\tau_a$ for each of the three source configurations.  The $\tau_d$ values were obtained from the measured pump bandwidth (shown in the first column); note that while $\tau_d$ tends to infinity for a monochromatic pump, for our realistic CW pump the two photon state is an incoherent sum of states, each associated with a monochromatic pump corresponding to each of the available spectral components, leaving $\tau_d$ undefined.  $\tau_a$ is obtained from the HOM dip width, specifically through a fit of the interferogram obtained in the case of the Gaussian model, equation (\ref{Eq:Cfortwolayers}), to the experimental data.

In accordance with our theory, the coincidence interferogram obtained for SPDC photon pairs pumped by a CW laser, see Fig. \ref{fig:QOCTresults}(a), presents an intermediate structure which can be either a dip or a peak with an amplitude proportional to  $\mbox{cos}(\omega_0 T)$ [see equation (\ref{E:CWintstru})].   For the specific case shown, this intermediate structure is a dip with a visibility larger than for the other two dips, i.e. those which correspond individually to the two layers. Figure \ref{fig:QOCTresults}(b) shows the QOCT interferogram obtained for partially anti-correlated photon pairs, i.e. with a pulsed pump in the ps regime.   Finally, Fig. \ref{fig:QOCTresults}(c) shows the QOCT interferogram obtained for nearly-factorable photon pairs,  i.e. with a pulsed pump in the fs regime.   These results show that the intermediate structure is reduced in amplitude for the ps pulsed pump case, and fully suppressed for the fs pulsed pump case, as predicted by our theory. Note that the separation of the two HOM dips, determined by the sample width,  is the same in all three configurations, and that the HOM dip width (depending on the anti-diagonal entanglement time $\tau_a$ and in our case determined by the filter spectral width), likewise remains essentially unchanged in all three configurations.   The dip separation corresponds well to $2 n L \approx 3$ mm (the factor of $2$ comes from the double pass  of the probe photon through the sample), with $n = 1.507$ the index of refraction of soda lime glass at $1550$ nm and $L \approx 1$ mm.  Note that the dip visibility nearly reaches the ideal value of $0.5$ for the CW and ps pump cases, while it is somewhat less for the fs case; this is probably due to greater instabilities in our fs Ti:sapphire oscillator.

\begin{figure}[htbp]
\centering
\includegraphics[width=12cm]{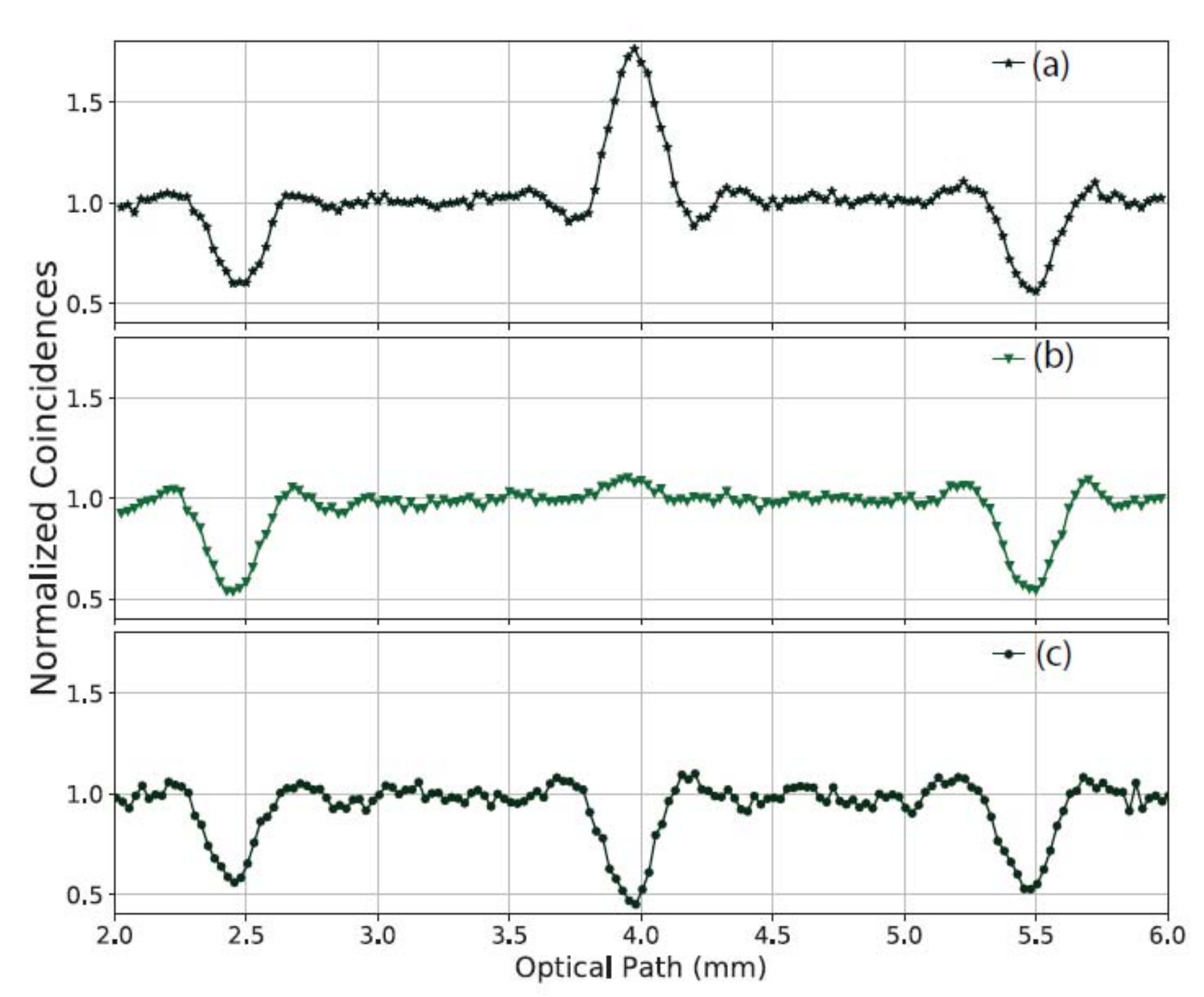}
\caption{(Color online). QOCT measurements of a two-layer sample, for a CW pump at a choice of three different frequencies highlighting the peak/dip behavior, corresponding to $\lambda_{p}=774.86$ nm (panel a),  $\lambda_{p}=774.94$ nm (panel b), and  $\lambda_{p}=775.50$ nm (panel c).}
\label{fig:MIRApicos}
\end{figure}

\begin{figure}[htbp]
\centering
\includegraphics[width=12cm]{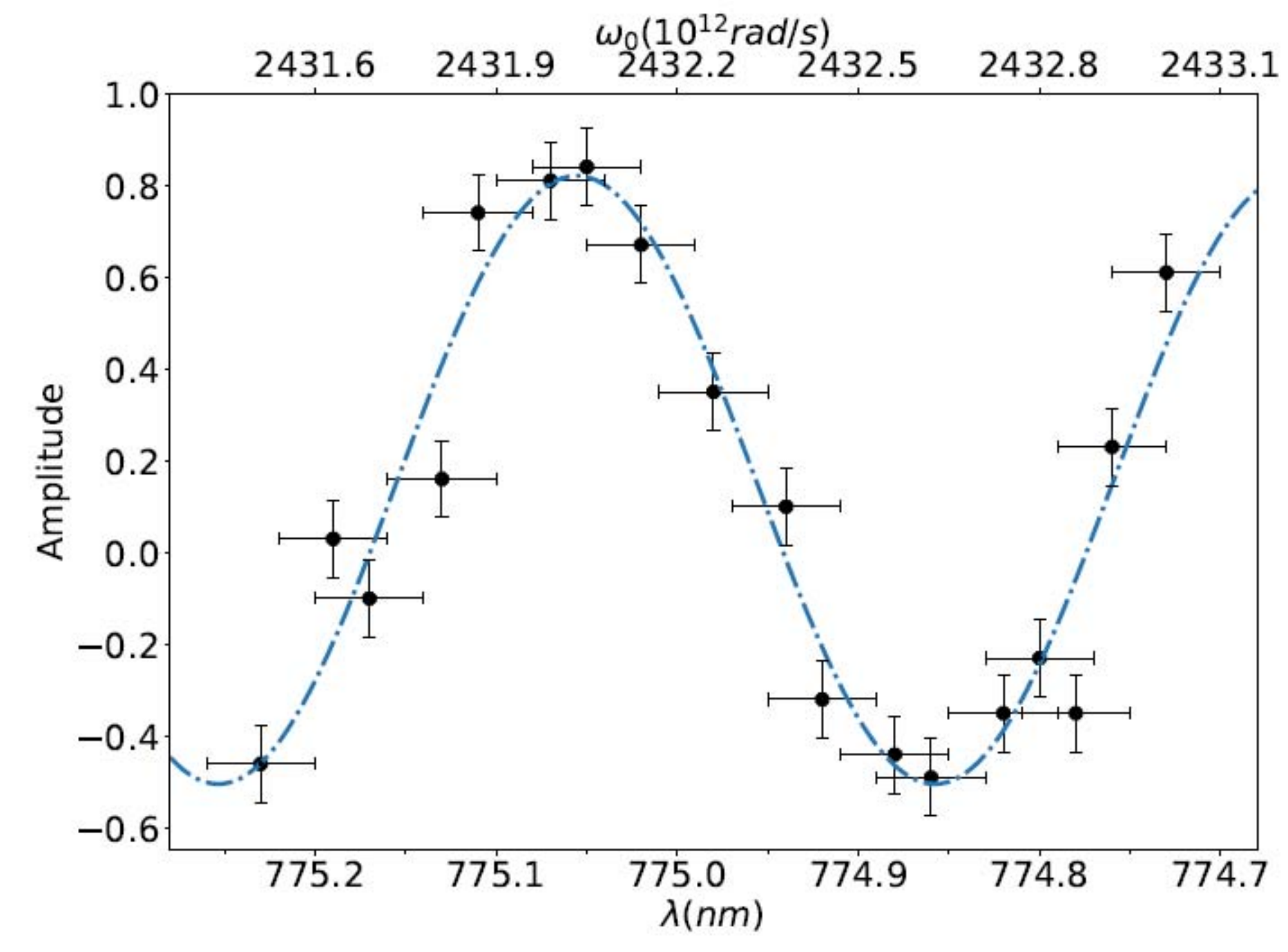}
\caption{(Color online) Visibility of the intermediate structure as a function of the pump frequency, for a CW pump in the SPDC process. The cosine behavior as expected from  equation (\ref{Eq:visibilities}) is apparent; we have shown a fit to the function $a\cos(T\omega_0+b)+c$, from which we obtain  $T \approx 10$ ps, as expected (the actual value is $T=10.067\pm0.419$ ps, calculated from the measurements provided in Table \ref{Tab:3tecnicas}).  \label{fig:QOCT_COSENO_ARTICULO}}
\end{figure}

We have pointed out that the intermediate structure can be either a dip or a peak, with an amplitude which depends on $\cos(\omega_{0}T)$. In order to demonstrate this effect experimentally,  we have carried out a measurement of the QOCT interferogram for the CW pump case, similar to that shown in Fig. \ref{fig:QOCTresults}(a),   as a function of the  pump frequency.    The full interferogram showing this dip/peak behavior is shown in Fig. \ref{fig:MIRApicos} for three specific choices of pump frequency.  The effect is further clarified in Fig. \ref{fig:QOCT_COSENO_ARTICULO}, which shows the amplitude of the intermediate structure as a function of the pump frequency obtained experimentally roughly within  one oscillation period, together with a sinusoidal curve obtained as a best fit to the data; the value of $T$ inferred from this data is about $10$ ps which corresponds well to the expected value $2 n L$ already mentioned and which also defines the separation between the two dips. Note that due to experimental imperfections the maximum amplitude of the peak in  Fig. \ref{fig:QOCT_COSENO_ARTICULO} is greater than the maximum amplitude of the dip.  It is worth mentioning that this peak/dip effect has been commented in previous works, see Abouraddy \emph{et al.} \cite{Abouraddy2002}, along with the suggestion that the intermediate structure could be suppressed by averaging over interferograms taken with multiple pump frequencies.   While it is questionable that an approach requiring multiple repetitions of the experiment, each with a slightly detuned pump frequency, is feasible in a practical setting,  our fs pulsed pump, which contains a range of pump frequencies, accomplishes such averaging process directly  leading to the suppression of the intermediate structure \emph{using a single experimental measurement}.

\begin{figure}[htbp]
\centering
\includegraphics[width=12cm]{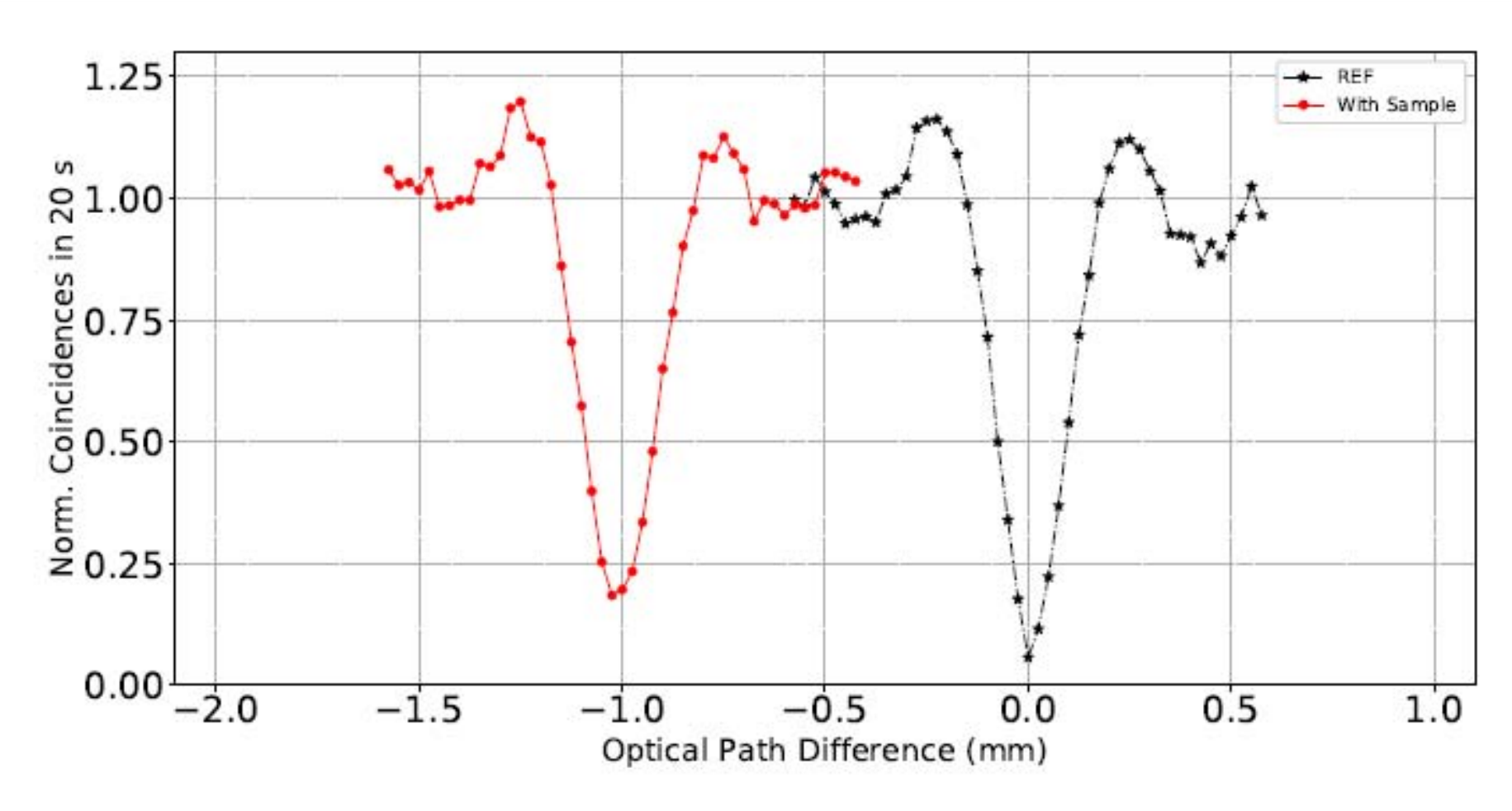}
\caption{(Color online) HOM dip displacement produced by transmission if the probe photon through our sample ($1$mm thickness of soda lime glass).  The displacement is proportional to $2(n-1)L$.  The curve composed of black stars corresponds to the the HOM obtained without sample, and the curve composed of red points corresponds to  the HOM obtained in the presence of the sample.}
\label{fig:HOM_ESPEJO_REF}
\end{figure}

Finally, we have implemented a complementary method designed to measure the thickness of a two-layer sample, where the two layers correspond to the air-sample and sample-air interfaces. In this method, we first utilize the same setup as described above (see Fig. \ref{fig:setup}), except that the sample is replaced by a mirror, resulting in a standard (single) HOM dip.   The experiment is now repeated with the sample (identical to the one used in the QOCT measurements)  introduced,  so that the photon now traverses the sample, is reflected from the mirror, and traverses the sample again upon reflection from the mirror.  This results in a translation of the HOM dip by time $2 (n-1) L/c$, which corresponds to the added optical path introduced by the sample.     We have shown our results for these measurements in Fig. \ref{fig:HOM_ESPEJO_REF}. Note that for the HOM obtained in the presence of the sample, the visibility is expected to be reduced (as is indeed the case in our experimental results), since the sample and mirror system behave as a three-layer QOCT sample with the visibility from each layer diminished in accordance with the retained flux from the layer in question.

We thus have at our disposal three  methods, see Table \ref{Tab:3tecnicas},  which provides combined information about the refractive index $n$  and the sample width $L$.  In two of these methods: i) based on the separation of the two HOM dips obtained int the QOCT measurement, and ii) based on fitting the intermediate structure amplitude as a function of pump frequency  to a sinusoidal function, we obtain the quantity $2 n L$.   In a third method, iii) based on measuring the HOM dip with and without sample, and monitoring the relative displacement, we obtain the quantity $2 (n-1) L$.   Therefore, we can use the result from any of the two techniques in the first group, together with the result of the third technique, so as to obtain two equations with two unknowns, i.e. $n$ and $L$.  Solving this set of equations yields an experimental measurement of $n$ and $L$ independently, without any \emph{a priori} knowledge about the sample.  We have selected method i), rather than method ii), because it entails a significantly lower uncertainty (see table).   These measurements  yield the values for $L$ and $n$ which are shown on the table, corresponding well with the nominal values from the manufacturer of the microscope slide.

\begin{table}
\caption[Table caption text]{Experimental determination of $n$ and $L$.}
\centering
\begin{tabular}{c c c c c }
   Technique & Measurement (mm) &  &  $n$ & $L$ (mm) \\
\hline \hline
\\
  $\begin{tabular}{*{1}{c}}
    QOCT dip separation \\  HOM with/without sample
  \end{tabular}$ & $\begin{tabular}{r}
    $2nL=3.021 \pm 0.025$ \\  $2(n-1)L=1.026 \pm 0.025$
  \end{tabular}$  & $\left. \begin{tabular}{c}
     \\  \\ 
  \end{tabular} \right\}$  & $1.514 \pm 0.025$ & $0.997 \pm 0.025$  \\
   \\
  cosine term (intermediate structure) & $2nL=3.033 \pm 0.072$ &  &  & \\
   
\end{tabular} 
\label{Tab:3tecnicas}
\end{table}

Our work offers interesting insights which could guide practical  implementations of  QOCT.   On the one hand, in a multilayer sample, a dip will occur for each interface, as well as an intermediate structure for each pair of layers. In the presence of many layers it becomes challenging to identify  which features in the interferogram signify the presence of a layer and which are spurious, i.e. due to the cross-correlation associated with the various pairs of layers in the sample:   if the experimenter is able to tune the pump frequency, those features which alternate between peak and dip could be identified as spurious.  Averaging over different experimental runs, each with a different value of the pump frequency, would then lead to the isolation of the dips which are attributable to a single layer.    A simpler solution suggested by our work  is to use a femtosecond pulsed pump for which the intermediate structures due to cross-correlation are suppressed entirely, resulting in the ability to obtain the desired metrological information about the sample \emph{with a single measurement} (i.e. with a single scan of the delay).  On the other hand, as we have discussed above, in a multi-layer sample the visibility of each HOM dip is reduced as $1/N$ where $N$ is the number of layers.  This means that depending on the level of noise, it may be difficult to obtain a good-quality measurement for samples which surpass a certain threshold in the number of layers.   In this context, the fact that the amplitude of the intermediate structures is unaffected by the number of layers means that despite the presence of many layers, information about the layer locations can be obtained from the intermediate peaks, despite the presence of an arbitrarily large number of layers.

\section*{Conclusions}

We have presented a theoretical study of quantum optical coherence tomography using photon pairs with arbitrary spectral correlations.  We have arrived at an expression in closed form for the QOCT interferogram, for the particular case of a two-layer sample with the photon pairs described through Gaussian model for the joint spectrum.   In this expression, it is clear that the interferogram includes a HOM dip corresponding individually to each sample layer (involving the probe photon being fully reflected by each layer),  in addition to a cross-correlation intermediate structure which corresponds to the probe photon being reflected partially from each of the two layers.   The intermediate structure can be either a dip or a peak depending on the sign of $\mbox{cos}(\omega_p T/2)$, where  $\omega_p$ is the pump frequency and $T$ is the propagation time between the two layers.  We have shown that as the pump bandwidth is increased, the amplitude of the intermediate structure is reduced, and for a sufficiently large pump bandwidth (for example, obtained for a fs pulsed pump in the SPDC process) the amplitude of the intermediate structure can be entirely suppressed.     We have discussed that in general for $N$ layers, the visibility of the dip associated with each layer is reduced as $1/N$ while the amplitude of the intermediate structure remains unaffected.  

We have presented an implementation of QOCT based on photon pairs in the telecommunications band produced by a PPLN crystal pumped by a Ti:Sapphire laser.   We have 
used three different configurations corresponding to our pump laser operated in the following modes:  i) continuous wave, ii) pulsed ps regime, and iii) pulsed fs regime.
For  each configuration we have measured the joint spectrum (albeit with some resolution limitations as discussed above), as well as the QOCT interferogram. We have shown explicitly that the amplitude of the intermediate structure is suppressed for our fs pulsed pump.    For the case of the CW pump, we have additionally shown that tuning the pump frequency leads to an alternating dip/peak behavior for the intermediate structure.  We have  performed a complementary experiment based on a HOM dip involving the probe photon being transmitted / not transmitted through the sample, so that the resulting dip displacement together with either the dip separation in the QOCT interferogram or a determination of the period of the dip/peak oscillations of the intermediate structure amplitude as a function of the the pump frequency, can be used so as to obtain an experimental measurement of the sample width and refractive index, independently, without \emph{a priori} knowledge of the sample.  We believe that our results may be a useful guide for future practical implementations of QOCT.

\section*{Methods}

\subsection*{Theoretical description of the coincidence interferogram}

Consider a $50:50$ beam splitter with input ports described by the operators $\hat{a}(t),\hat{b}(t)$, and output ports described by $\hat{c}(t),\hat{d}(t)$. The input and output operators are related by
\begin{equation}
\hat{c}(t) =  \frac{1}{\sqrt{2}} \int \mathrm{d} \omega e^{i \omega t}  \left[i\hat{a}(\omega)  + \hat{b}(\omega)\right], \quad  \hat{d}(t) = \frac{1}{\sqrt{2}} \int \mathrm{d} \omega e^{i \omega t} \left[\hat{a}(\omega) + i\hat{b}(\omega)\right].
\end{equation}
The coincidence rate is given by the correlation $\langle \hat{n}_{c}(t_{1}) \hat{n}_{d} (t_{2}) \rangle = \langle \hat{c}^{\dagger}(t_{1})\hat{c}(t_{1}) \hat{d}^{\dagger}(t_{2})\hat{d}(t_{2})  \rangle $.  Introducing a delay $\tau$ in the input port $\hat{a}$ and the sample $H(\omega)$ in $\hat{b}$, the output operators are written as
\begin{equation}
\hat{c}(t) =  \frac{1}{\sqrt{2}} \int \mathrm{d} \omega e^{i \omega t}  \left[i\hat{a}(\omega)e^{i\omega \tau}  + H(\omega)\hat{b}(\omega)\right], \quad  \hat{d}(t) = \frac{1}{\sqrt{2}} \int \mathrm{d} \omega e^{i \omega t} \left[\hat{a}(\omega)e^{i\omega \tau} + i H(\omega) \hat{b}(\omega)\right].
\end{equation}
Consider the input state of the form 
\begin{equation}
|\psi \rangle = \int \int \mathrm{d} \omega'_{1} \mathrm{d} \omega'_{2} f(\omega'_{1},\omega'_{2}) \hat{a}^{\dagger}(\omega'_{1})\hat{b}^{\dagger}(\omega'_{2}) |0\rangle.
\end{equation}
Then, in order to get $\langle \hat{n}_{c}(t_{1}) \hat{n}_{d} (t_{2}) \rangle$ we start with 
\begin{align}
\hat{c}(t_{1})\hat{d}(t_{2})|\psi\rangle & = \frac{1}{2}\int \int \int \int \mathrm{d}\omega_{1}\mathrm{d}\omega_{2}\mathrm{d}\omega'_{1}\mathrm{d}\omega'_{2} e^{i\omega_{1}t_{1}} e^{i\omega_{2}t_{2}} \times \\ 
& \left[ i \hat{a}(\omega_{1})e^{i\omega_{1}\tau} + H(\omega_{1})\hat{b}(\omega_{1})   \right] \left[ \hat{a}(\omega_{2})e^{i\omega_{2}\tau} + i H(\omega_{2})\hat{b}(\omega_{2})   \right] f(\omega'_{1},\omega'_{2}) \hat{a}^{\dagger}(\omega'_{1})\hat{b}^{\dagger}(\omega'_{2}) |0\rangle.
\end{align}
By expanding and using 
\begin{align}
\hat{a}(\omega_{1})\hat{a}(\omega_{2}) \hat{a}^{\dagger}(\omega'_{1})\hat{b}^{\dagger}(\omega'_{2}) |0\rangle &= 0, \\
\hat{b}(\omega_{1})\hat{b}(\omega_{2}) \hat{a}^{\dagger}(\omega'_{1})\hat{b}^{\dagger}(\omega'_{2}) |0\rangle &= 0, \\
\hat{a}(\omega_{1})\hat{b}(\omega_{2}) \hat{a}^{\dagger}(\omega'_{1})\hat{b}^{\dagger}(\omega'_{2}) |0\rangle &= \delta(\omega_{1}-\omega'_{1})\delta(\omega_{2} - \omega'_{2}),\\
\hat{b}(\omega_{1})\hat{a}(\omega_{2}) \hat{a}^{\dagger}(\omega'_{1})\hat{b}^{\dagger}(\omega'_{2}) |0\rangle &= \delta(\omega_{1}-\omega'_{2})\delta(\omega_{2} - \omega'_{1}),
\end{align}
we obtain 
\begin{align}
\hat{c}(t_{1})\hat{d}(t_{2})|\psi\rangle & = \frac{1}{2}\int \int \int \int \mathrm{d}\omega_{1}\mathrm{d}\omega_{2}\mathrm{d}\omega'_{1}\mathrm{d}\omega'_{2} e^{i\omega_{1}t_{1}} e^{i\omega_{2}t_{2}} \times \\ 
& \left[   -\delta(\omega_{1}-\omega'_{1})\delta(\omega_{2}-\omega'_{2})e^{i\omega_{1}\tau}H(\omega_{2}) + \delta(\omega_{1}-\omega'_{2})\delta(\omega_{2}-\omega'_{1})e^{i\omega_{2}\tau}H(\omega_{1}) \right] f(\omega'_{1},\omega'_{2}) |0\rangle.
\end{align}
Integrating over $\omega'_{1},\omega'_{2}$ results in
\begin{equation}
\hat{c}(t_{1})\hat{d}(t_{2})|\psi\rangle  = \frac{1}{2}\int \int \mathrm{d}\omega_{1}\mathrm{d}\omega_{2} e^{i\omega_{1}t_{1}} e^{i\omega_{2}t_{2}}   \left[-e^{i\omega_{1}\tau}H(\omega_{2})f(\omega_{1},\omega_{2}) + e^{i\omega_{2}\tau}H(\omega_{1})f(\omega_{2},\omega_{1}) \right]  |0\rangle. \label{Eq:procedure1}
\end{equation}
It follows that $\langle \hat{n}_{c}(t_{1}) \hat{n}_{d} (t_{2}) \rangle = \langle \hat{c}^{\dagger}(t_{1})\hat{c}(t_{1}) \hat{d}^{\dagger}(t_{2})\hat{d}(t_{2})  \rangle = \langle \hat{c}^{\dagger}(t_{1}) \hat{d}^{\dagger}(t_{2})\hat{c}(t_{1})\hat{d}(t_{2})  \rangle $, and therefore, by using equation (\ref{Eq:procedure1}) we have that
\begin{align}
& \langle \hat{n}_{c}(t_{1}) \hat{n}_{d} (t_{2}) \rangle = \frac{1}{4} \int \int \int \int \mathrm{d}\omega_{1}\mathrm{d}\omega_{2}\mathrm{d}\omega'_{1}\mathrm{d}\omega'_{2} e^{i\omega_{1}t_{1}} e^{i\omega_{2}t_{2}} e^{i\omega'_{1}t_{1}} e^{i\omega'_{2}t_{2}} \times \nonumber \\
&  \left[-e^{i\omega_{1}\tau}H(\omega_{2})f(\omega_{1},\omega_{2}) + e^{i\omega_{2}\tau}H(\omega_{1})f(\omega_{2},\omega_{1}) \right]  \left[-e^{-i\omega'_{1}\tau}H^{\ast}(\omega_{2})f^{\ast}(\omega_{1},\omega_{2}) + e^{-i\omega_{2}\tau}H^{\ast}(\omega_{1})f^{\ast}(\omega_{2},\omega_{1}) \right]. \label{Eq: procedure2}
\end{align}

The coincidence rate $C(\tau)$ is equal to the time average of the previous equation. We use the integrals
\begin{equation}
\frac{\omega_{1}-\omega'_{1}}{2\pi} \int_{0}^{\frac{2\pi}{\omega_{1}-\omega'_{1}}}    \mathrm{d}t_{j} e^{i(\omega_{j}-\omega'_{j})t_{j}} = \delta(\omega_{j} - \omega'_{j}), \qquad j=1,2,
\end{equation}
to evaluate the time average of equation (\ref{Eq: procedure2}). The result is
\begin{equation}
C(\tau) = \frac{1}{4} \int \int \mathrm{d}\omega_{1}\mathrm{d}\omega_{2} | -e^{i\omega_{1}\tau}H(\omega_{2})f(\omega_{1},\omega_{2}) + e^{i\omega_{2}\tau}H(\omega_{1})f(\omega_{2},\omega_{1})  |^{2}.
\end{equation}
Finally, we factor out $e^{i\omega_{1}\tau}$ to get the result used in equation (\ref{Eq:coinc_interf}):
\begin{equation}
C(\tau) = \frac{1}{4} \int \int \mathrm{d}\omega_{1}\mathrm{d}\omega_{2} \left| -H(\omega_{2})f(\omega_{1},\omega_{2}) + e^{i(\omega_{2}-\omega_{1})\tau}H(\omega_{1})f(\omega_{2},\omega_{1})  \right|^{2}.
\end{equation}

\subsection*{Details of the experiment}

The experiment (see Fig. \ref{fig:setup}) is mainly composed of three parts: (1) the pump system and the source of SPDC photons, (2) the HOM interferometer, and (3) the detection system.

\textbf{The pump system.} We used two different pump sources, our first pump source was a laser cavity (a Ti:Sapph laser built in our lab) that produces femtosecond pulses with a repetition rate of $90$ MHz with an average power of $400$ mW. The central wavelength of the pulse is tuned by using a slit after a dispersion-compensating prism. In this way, we achieve a tunable range between $750$ nm and $800$ nm, with bandwidths about $10$ to $20$ nm, approximately. The beam waist has a radius of $415$ $\mu$m at the $e^{-1}$ amplitude value. For our measurements, we use a $775$ nm central wavelength with $14$ nm bandwidth and attenuate the beam power to $10$ mW using neutral density filters.

Our second pump laser is a picosecond laser (COHERENT Mira 800) which can operate in continuous mode (CW)($\delta\lambda\approx 0.15$nm) or pulsed mode (PW) ($\delta\lambda\approx 15$nm). In both cases we set the central frequency to $775$nm.

\textbf{The source of SPDC photons [shown in Fig. \ref{fig:setup} (part 1)].} We use a MgO-doped PPLN nonlinear crystal with dimensions $0.5 \times 10 \times 10$ mm (thickness, width, length). The periodicity of the poling goes from $18.50$ to $20.90$ $\mu$m with steps of $0.3$ $\mu$m. The period is $0.5$mm. The crystal can generate photon pairs through the process of SPDC type 0 (eee,ooo), which means, both the pump photon and photon pairs have the same polarization. We can tune the phase matching condition of the crystal by controlling  its temperature. With the PPLN oven we can adjust the crystal temperature in the range of $30^{\circ}$ to $200^{\circ}$ using a temperature controller (TC), that give us a range of wavelengths between $1520$ and $1620$ nm. We produce degenerate photon pairs at $1550$nm using a poling period of $19.10$ $\mu$m and a temperature of $90^{\circ}$. To achieve high yield of photon pairs, we focus the pump laser into the crystal by using a plano-convex spherical lens with antireflection coating (L) with a focal lens of $15$mm. We achieve a focused beam waist of $46$ $\mu$m with a Rayleigh range of $11.2$mm. The photon pairs are emitted forming a cone with half-angle of $1.2^{\circ}$. We use a set of filters (SF), it includes a high-pass filter with a cut-off wavelength of $980$nm to eliminate the pump after the crystal and a bandpass filter ($1550\pm 1.5$ nm) to ensure the spectral quality of the photon pairs.

\textbf{The detection system [shown in Fig. \ref{fig:setup} (part 3)].} We use a diode laser at $1550$ nm to align the lenses for coupling the photon pairs into PANDA-type polarization-maintaining single-mode fibres. We position the lenses according to the calculated angle of emission of the photon pairs, which is $2.4^{\circ}$ for the cone aperture. The distance from the crystal was optimized according to the theory found in Vicent \emph{et al.} \cite{Vicent2010} We use aspheric lenses with $15$ mm focal length, mounted on $3$-axis linear translation stages with micrometer resolution (Thorlabs MicroBlocks). The single-mode fibres are connected to fibre-coupled APD photodetectors (id-230 from id-quantique). These detectors operate in free-running mode, at a temperature of $-90^{\circ}$ C, and allow the control of detection dead-time and quantum efficiency. We use a dead-time of $10\mu$s and quantum efficiency of $15\%$. Dark counts are in the order of dozens of counts per second for signals with thousands of counts per second.

\textbf{The HOM interferometer [shown in Fig. \ref{fig:setup} (part 2)].} After coupling the photon pairs to the fibres, each of them passes through a manual fibre polarization controller (MPC1). The probe photon goes to a polarization maintaining optical circulator (PMC) that sends the photon to the sample (S) and directs the reflected photon to one input of a $2 \times 2$ fibre beam splitter (BS). We remark that all fibre elements maintain polarization. The reference photon first passes through a compensating fibre (FC) that matches the fibre length of the circulator, and then through a free-space delay that consists of a pair of coupling stages. The distance to the second coupling stage is controlled by a linear translation stage(HOM stage). This translation stage provides the delay $\tau$ in the theory. Then, the fibre is connected to another manual manual fibre polarization controller (MPC2) to match the polarization of probe and reference photons at the fibre beam splitter (BS). The outputs of the fibre beam splitter are sent to Avalanche Photodetectors (APD1 and APD2), which are connected to channels 1 and 2 of a time tagger (TDC id800) to measure coincidence events. For the initial alignment, we start by measuring a HOM dip by using a mirror (RM) without the sample (S). We make sure that the lengths of the fibre in both arms of the interferometer are balanced by fusing together optical fibres.  We swept the translation stage and measure the coincidence interferogram until we find the HOM dip. The location of the HOM dip is our reference to start scanning the sample.

\bibliography{biblio}

\section*{Acknowledgements}
Dorilian Lopez-Mago acknowledges support from Consejo Nacional de Ciencia y Tecnolog\'{i}a (CONACYT) through the grants 257517, 280181 and 293471. A.B.U. acknowledges support from PAPIIT (UNAM) grant IN104418, CONACYT Fronteras de la Ciencia grant 1667, and AFOSR grant FA9550-16-1-1458.

\section*{Author contributions statement}
D.L.M. and A.B.U. conceived the idea and the experiment. P.Y.G., A.M.A.M., and H.C.R. conducted the experiments. G.C.O., M.R.A, J.G.M., and R.R.A  built the femtosecond laser used in the experiment. D.L.M., A.B.U., P.Y.G. and A.M.A.M. analysed the results and reviewed the manuscript. 

\section*{Additional information}
\textbf{Competing interests:} The authors declare no competing interests.

\end{document}